\documentclass[a4paper,12pt]{article}
\usepackage[cmex10]{amsmath}
\usepackage{latexsym,amssymb}
\usepackage[mathscr]{eucal}
\usepackage{amsxtra,amscd,amsthm,upref,layout,color}
\usepackage{calc}
\usepackage[final]{graphicx}
\def\ac{\hfill\break\noindent }
\def\refup#1{{$^{#1}$}}
\def\p{^{^{\prime}}}

\def\barn{{\bar n}}
\def\br{{\bf r}}\def\bq{{\bf q}}\def\rdv{\rm{dv}}\def\rd{\rm{d}}
\def\rv{{\rm v}}
\def\oo{{\leavevmode\setbox0=\hbox{h}\dimen0=\ht0 \advance\dimen0
by-1ex\rlap{\raise0.47\dimen0\hbox{\char'27}}o}}
\def\begeq{\begin{equation}}
\def\endeq{\end{equation}}
\def\begdis{\begin{displaymath}}
\def\enddis{\end{displaymath}}

\def\cC{{\cal C}}
  \def\cI{{\cal I}}
  
\def\cN{{\cal N}}  
\def\cS{{\cal S}}

\def\ie{{\em i.e.}}%\def\ref#1{$^{#1}$}
\def\etal{{\em et al.}}
\def\hw{{\hat \omega}}

\def\tn{{\tilde {n}}}

\def\teta{{\tilde {\eta}}}

\def\Iv{{I_V}}\def\Vo{{V_{O}}}\def\Ivo{{I_{\Vo}}}\def\cIv{{{\cI}_V}}\def\cIvo{{{\cI}_{\Vo}}}

\begin{document}
\title{Scattering intensity limit value at very small angles}
\author{ %1
{{Salvino Ciccariello}}\\
%%%%%%%%
%%       \begin{minipage}[t]{0.9\textwidth}
%%       \begin{flushleft}
%%       \setlength{\baselineskip}{12pt}
%%    {\slshape  {\footnotesize{Universit\`{a} di Padova,
%%       Dipartimento di Fisica {\em G. Galilei}}}},\\
%%      {\slshape{\footnotesize{Via Marzolo 8, I-35131 Padova, Italy}} }\\
%%      \footnotesize{salvino.ciccariello@unipd.it}
%%    \end{flushleft}
%%    \end{minipage}
%\\[10mm]
%}      %1
%
%%%%%%%%
  \begin{minipage}[t]{0.9\textwidth}
   \begin{flushleft}
\setlength{\baselineskip}{12pt}
{\slshape  {\footnotesize{Universit\`{a}  {\em {C\`{a} Foscari}}, Department of Molecular Sciences and Nanosystems, 
 Via Torino 155/B, I-30172 Venezia, Italy, and \\
Universit\`{a} di Padova,
Dipartimento di Fisica {\em G. Galilei}, 
Via Marzolo 8, I-35131 Padova, Italy.}}}\\
 \footnotesize{salvino.ciccariello@unipd.it}
\end{flushleft}
\end{minipage}
%\\[10mm]
}      %1
%\\[10mm]
 \date{\today}
     % Use \shortauthor to indicate an abbreviated author list for use in
     % running heads (you will need to uncomment it).
%
%\\[10mm]
\date{\today}
     % Use \shortauthor to indicate an abbreviated author list for use in
     % running heads (you will need to uncomment it).
\maketitle                        % DO NOT DELETE THIS LINE
\begin{abstract} \noindent 
The existence of the limit of  a  sample scattering intensity, as the scattering 
vector approaches  zero, requires and is ensured by the property that the 
mean value of the scattering density fluctuation over  volume $V$   
asymptotically behaves,  at large $V$s, as $\nu V^{-1/2}$, $\nu$ being an 
appropriate constant.  Then, the  
limit of the normalized scattering intensity is equal to $\nu^2$. The 
implications of this result are also analyzed  in the case 
of samples made up of two homogeneous phases.  \\

\noindent Synopsis: {\em  The mean value of the scattering density 
fluctuation must asymptotically behave as $V^{-1/2}$ for the scattering 
intensity limit at  reciprocal space origin to exist.}\\

% 1st draft July 4, 2016\\
% 2nd draft: Aug. 8, 2016. 
\end{abstract}
%\begin{synopsis}
%................
%\end{synopsis}
\vfill
%\rightline{DFPD 2011/CM/9}
\eject
{}{}
%.............................................................................80
%%%%%%%%%%%         INTRODUCTION
\subsubsection*{Introduction}  
The aim of this note is to  discuss the existence as well as the meaning 
of $I(0^+)$, \ie\ the limit  of $I(\bq)$, the elastic scattering intensity of X-rays 
or neutrons, as the modulus ($q$) of the scattering vector ($\bq$) 
approaches zero.   It is well known that $I(0^+)$ is linearly related to the 
isothermal compressibility of the sample if this is a fluid  made up of 
identical particles [Guinier \& Fournet (1955), Hansen \& MacDonald 
(1976)]. Oppositely, to the author knowledge, no general expression  was 
known in the case of sample made up of two homogeneous 
phases   (as it is commonly assumed in the small-angle scattering realm) till 
a recent paper by Gommes (2006) who showed that $I(0^+)$ is equal to the 
variance of the scattering density fluctuation under the assumption that this 
consists of a collection of  independent  and equally distributed random variables  (Rosenthal,  2000). \ac 
Aim of  this note is to point out a different interpretation of  $I(0^+)$, namely:   
the integral of a physical scattering density fluctuation over a volume $V$,  at large $V$s, asymptotically behaves as $\nu\,V^{1/2}$  where 
the numerical coefficient $\nu$ is  related to the  $I(0^+)$ value by the simple 
relation $I(0^+)=\nu^2$. \ac 
To begin with, it is convenient first to recall a basic assumption
usually understood in dealing with the scattering experiment results from a 
matter sample:  no change is experimentally observable if one varies the volume $V$ as well as the center of gravity position $O$ of the sample's illuminated portion $V_{O}$ with respect to the ingoing beam and/or  if one cuts another sample from a given material specimen. [For a discussion of these aspects 
in the case of stereological analysis one should refer to a  report by 
Lantuejoul (1990).]  The expression of  the scattering intensity $\cIvo(\bq)$, relevant  
to a sample illuminated volume $\Vo$,   is simply given 
by [Guinier \& Fournet (1955), Kostorz (1979), Feigin \& Svergun (1987)]
\begeq\label{1.5A}
\cIvo(\bq)=|{\tn}_\Vo(\bq)|^2,
\endeq 
where ${\tn}_\Vo(\bq)$ denotes the Fourier transform (FT) of $n_\Vo(\br)$, the 
scattering density of the illuminated portion of the sample. This 
function   is defined as being equal to $n(\br)$ (the scattering density of the 
infinitely large sample) if the tip of $\br$ falls inside $\Vo$, the sample illuminated part having its gravity center set at point $O$, and to zero elsewhere. It is also recalled that 
the $n(\br)$ determination requires a  statistical mechanical average or the use 
of the functional density theory  since the only small angle scattering intensity is 
not sufficient for its determination.   Even 
though a Dirac $\delta(\cdot)$-like approximation 
of $n(\br)$ is sometimes adopted  in the case of perfect crystals,  it will  
be assumed, in the first part of this note, that $|n(\br)|$ is a continuous bounded function and, in the second, that it is a discrete valued function. 
The $n(\br)$ and $n_V(\br)$ units  are $L^{-3}$ and $L^{-2}$ 
in the case of X-ray and neutron scattering, respectively. Moreover, the 
$\cIvo(\bq)$ definition refers to an ingoing beam of unit intensity and it also understands that the electron  Thomson  factor, present  in the only 
case of X-ray scattering, be factorized  out. We explicitly restate now the two assumptions on which the following analysis rests:
\begin{itemize}
\item A) the scattering density is either a continuous or a discrete valued 
function, in both cases with lower and upper finite bounds. This assumption 
clearly confines the scattering vector to the small-angle domain and excludes  
fractal systems from our analysis;
\item B) once $V$ is larger or of the size considered in scattering experiments, 
the {\em observed} scattering intensity per unit volume $\Ivo(\bq)=\cIvo(\bq)/V$ 
is independent on $V$ and on $O$, which justifies the standard notations $I(\bq)$ and $\cI(\bq)$.
\end{itemize}
%%%%%%%%  \subsubsection*{The continuous scattering density case}     %%%%
\subsubsection*{The continuous scattering density case}
%%%%%%%%%%%                                        \end{document}
Equation (\ref{1.5A}) implies that  
\begeq\label{1.5B}
\cIvo(0)={{\bar n}_\Vo}^2\,V^2,
\endeq
where ${\bar n}_\Vo$ denotes the mean value of $n(\br)$ over $\Vo$. 
Guinier \& Fournet(1955) (hereafter referred to as I)  
already  stressed that this mean value differs from ${\bar n}[\equiv \lim_{_{V\to \infty}}(\int_V n(\br)\rdv /V)]$, the mean scattering density value of  the infinitely 
large sample.  Similarly to what reported in section 2 of Landau \& Lifshitz (1967a), let us assume now  that, as $\Vo$ gets larger and larger,  $ {\bar n}_{\Vo}$ asymptotically  behaves as 
\begeq\label{1.5C}
 {\bar n}_{\Vo}\approx {\bar n} + \nu_o/V^{1/2}+{\nu_o}\p/V^{\epsilon} 
\endeq
 with  $\epsilon_o>1$ and $\nu_0$ and ${\nu_o}\p$ suitable constants, 
eventually depending on the position $O$. In this way, by (\ref{1.5B}), one would 
find that 
\begeq\label {1.5D}
\cIvo(0)\approx {{\bar n}}^2\,V^2 + 2{\bar n}\,\nu_o\, V^{3/2} + \nu_o^2\,V+\ldots
\endeq 
This relation shows that $\cIv(0)$, similarly to  the intensity 
values observed at the Laue spots in the case of crystalline samples (Landau \& Lifshitz, 1967b), scales as $V^2$ provided  $V$ is sufficiently large. But, 
in contrast with the intensity values at the Laue spots [different from $( 0,0,0)$],  
 the $\cIv(0)$ value is experimentally not observable due to the beam stop 
presence.  Hence, the value of the scattering intensity at the origin of reciprocal space can only be obtained extrapolating  the collected $\cIvo(\bq)$ values 
towards the origin. However, according to B),  for the $\bq$s lying outside the beam stop $\Ivo(\bq)$ behaves as an intensive quantity, also independent on $O$, once $V$ is not smaller than the size usually employed in experiments. Therefore, assuming that the aforesaid extrapolation procedure be unambiguous and denoting  the resulting value by $\cIvo(0^+)$,  the requirement that   the $O(V)$ contribution present  (\ref{1.5D}) continuously matches the extrapolated one yields 
\begeq \label{1.5.E}
\Ivo(0^+)= \cIvo(0^+)/V\approx \nu^2, 
\endeq
\ie\ the $q\to 0$  limit value of the scattering intensity per unit volume of the 
illuminated sample is determined by the coefficient of the $O(V^{-1/2})$ term 
present 
in the ${\bar n}_V$ asymptotic expansion. This, moreover,  must have the form reported in  (\ref{1.5D}) with $\nu$ independent  on $O$, as we already wrote in 
(\ref{1.5.E}). 
To make the above argument  rigorous one needs to show that $\cIv(q)$ 
shows up a peak that fully lies behind the beam stop and that the peak 
value, as $V$ increases, behaves according to (\ref{1.5D}).   This point 
is thoroughly explained in I, and we simply mention the main steps. 
In order to separate the peak contribution,  one introduces the so-called 
scattering density fluctuations $\eta(\br)$  of the infinitely large sample 
according to the definition 
\begeq\label{1.5F}
\eta(\br)\equiv n(\br)-\barn.
\endeq  
The scattering density fluctuation of the illuminated part of the sample will 
be denoted by $\eta_{\Vo}(\br)$ and, similarly to $n_{\Vo}(\br)$, it coincides with   
$\eta(\br)$ inside the illuminated portion of the sample and is equal to zero elsewhere so as to write 
\begeq\label{1.7A}
n_{\Vo}(\br)= \eta_{\Vo}(\br)+\barn\, \Theta_{\Vo}(\br),
\endeq
where $\Theta_{\Vo}(\br)$ is defined as being equal to 1 if the tip $\br$ falls 
inside the illuminated part ${\Vo}$ of the sample and to zero elsewhere.  By Fourier transforming (\ref{1.7A}),  substituting the result in (\ref{1.5A}) and recalling that $\cIvo(\bq)=V\Ivo(\bq)$ one gets 
\begeq\label{1.8A}
\Ivo(\bq)=\Big[\barn^2|{\tilde \Theta_{\Vo}}(\bq)|^2+2{\rm Re}\big(
{\tilde\Theta}_{\Vo}(\bq){\overline{{\teta}_{\Vo}}}(\bq)\big)\Big]/V+
|{\teta}_{\Vo}(\bq)|^2/V.
\endeq
[Here the large overbar denotes the complex conjugate and the tilde the FT.]  
As explained in I,  the contribution inside the square brackets is 
restricted to an angular range fully hidden by  the beam-stop once $V$ has 
approached a size two-three order of magnitudes smaller than that employed 
in typical experiments. Besides, it approaches to a Dirac $\delta$ function as 
$V\to \infty$. 
Hence, it does not contribute to the limit of the observed $\Ivo(\bq)$ as $q\to 0$   
and $\Ivo(0^+)$ is fully determined by the limit of the second term on the rhs of (\ref{1.8A}).  
In conclusion, the large value of $V$ makes it accurate to write, for 
$q>0$,  
\begeq\label{1.9A}
I_{\Vo}(\bq)=|{\teta}_{\Vo}(\bq)|^2/V=\int _{R^3}e^{i\bq\cdot\br}\gamma_{\Vo}(\br)\rdv,
\endeq 
where $\gamma_{\Vo}(\br)$,  the non-normalized correlation function of the sample, 
is defined as 
\begeq\label{1.9Ab}
\gamma_{\Vo}(\br)\equiv \frac{1}{V}\int\eta_{\Vo}(\br_1)\eta_{\Vo}(\br_1+\br)\rdv_1.
\endeq
The $\bq\to 0$ limit of  (\ref{1.9A}) yields 
\begin{eqnarray}
I_{\Vo}(0^+)&\approx& I(0^+)=\lim_{V\to \infty}|{\teta}_{\Vo}(0)|^2/V=
\lim_{V\to\infty}\bigl[\int_{R^3}\gamma_{\Vo}(\br)\rdv\big]=\nonumber \\
\quad & &\quad\quad\quad
\lim_{V\to \infty}\Big[
\int \eta_{\Vo}(\br)\rdv\Big/V^{1/2}\Big ]^2.\label{1.10A}
\end{eqnarray}  
By construction, the scattering density fluctuation definition (\ref{1.5F}) 
implies that its mean value $\bar\eta$ is equal to zero, \ie\,
${\bar\eta}\equiv\lim_{V\to\infty}(\int \eta_{\Vo}(\br)\rdv/V)=0$. It is noted that  this limit value 
does not imply that  $\lim_{V\to\infty}\int \eta_{\Vo}(\br)\rdv=0$. In fact, for the 
first limit to be valid,   it is necessary and sufficient that, at large $V$s, one 
asymptotically finds 
\begeq\label{1.10B}
\Big|\int \eta_{\Vo}(\br)\rdv\Big|\approx \big|\nu_o\big|\, V^{\alpha}\quad {\rm with}\quad \alpha<1, 
\endeq
and $\nu_o$ constant.  Substituting this behavior into the rightmost member  of 
(\ref{1.10A}) one finds that the quantity inside the square brackets behaves 
as ${\nu_o}^2\,V^{2\alpha-1}$. Since we are in a $V$ range where assumption B) applies, $\Iv(0^+)$ is intensive with respect to $V$ and, for this to happen, it must result  $2\alpha-1=0$, \ie\,$\alpha=1/2$.  
Then, equation (\ref{1.10B})   becomes  
\begeq\label{1.10Ca}
\Big|\int \eta_{\Vo}(\br)\rdv\Big|\approx |\nu|\, V^{1/2} ,
\endeq
where assumption B) again requires that $|\nu|$ does not depend on $O$ that 
was therefore omitted as suffix. In appendix A we show that the above equation 
coincides with 
\begeq\label{1.10C}
\int \eta_{\Vo}(\br)\rdv\approx \nu\, V^{1/2},
\endeq 
if one assumes that $\nu\ne 0$ and we also report an example 
of function obeying condition (\ref{1.10C}). 
Using definition (\ref{1.5F}), one immediately realizes that 
condition (\ref{1.10C})  coincides with a weakened form of  (\ref{1.5C}) in so 
far the condition $\epsilon>1$ is now substituted by $\epsilon>1/2$ because 
$\bar\eta=0$.  Besides, 
it is not necessary to assume the validity of the weakened  (\ref{1.5C}) because 
this condition is a consequence of the intensive nature of $\Ivo(0^+)$ with 
respect to $V$. The basic conclusion of this analysis follows from equations 
(\ref{1.10C}) and  (\ref{1.10A}). It states that: the $\bq\to 0$ limit of the 
scattering intensity is equal to the squared coefficient in front of the 
leading $O(V^{-1/2})$ term of the asymptotic expansion of  the mean value of the 
scattering density fluctuation as $V\to \infty$, \ie
\begeq\label{1.10D}
I_V(0^+)\approx I(0^+)=\nu^2.
\endeq 
This quantity is certainly positive and varies within the range $[0,\infty)$,  the 
outermost values being clearly assumed in the proximity of   possible critical 
points.  It is also noted that the $\nu^2$ units are $[L^{-3}]$ for X-ray scattering 
and $[L^{-1}]$ for neutron one.\\ 
From equation (\ref{1.10C}) it is possible to derive a further consequence of 
some interest, namely:  the angular average  of the scattering density fluctuation, 
at fixed distance $r$ from a point $O$, decreases as $r^{-3/2}$, 
whatever $O$, at very large $r$s.  
More precisely, one has  
\begeq\label{1.11}
\frac{1}{4\pi}\int \eta(r\hw)\rd\hw \approx \frac{\nu}{2\sqrt{3} (4\pi r)^{3/2}}.
\endeq
[Here $\hw$ denotes a unit vector that spans all possible directions.] 
To prove this relation, consider a spherical shell $\cS$ of  center $O$, thickness $\delta$  and inner radius $r$. Denote by $V_1$ and $V_2$ the spheres 
centered at $O$ and having radii 
equal to $r$ and $r+\delta$ and assume that the spheres are sufficiently 
large to make (\ref{1.10C}) valid. It results that 
\begeq\nonumber
\int_{\cS} \eta(\br)\rdv=\int\eta_{V_2}\rdv-\int\eta_{V_1}\rdv\approx 
\nu(V_2^{1/2}-V_1^{1/2})
\endeq 
The difference of the two integrals, evaluated up to terms $O(\delta)$, is 
equal to $4\pi r^2\delta \int \eta(r\hw)\rd \hw$, while 
$V_2^{1/2}-V_1^{1/2}\approx (4\pi r^3/3)^{1/2}\delta/2r$. By  these  
two expressions one immediately recovers result (\ref{1.11}).\\ 
Relation (\ref{1.10C}) is fully general and represents the basic result of this 
note. 
%%%%%%%
\subsubsection*{The discrete valued scattering density case}
%%%%%%%
We analyze now the  implications of (\ref{1.10C}) when one assumes that the scattering 
density fluctuation has the form pertinent to a two homogeneous phase 
sample [Debye \etal\,(1957), Ciccariello(2002)], the case most typically 
considered in the small-angle scattering realm. As customary, we denote 
by $\rho_1(\br)$ and $\rho_2(\br)$ the characteristic functions of phases 
1 and 2 that respectively have scattering density values equal to $n_1$ and 
$n_2$.   [We also recall that $\rho_1(\br)$ is, by definition, equal to 1 if the 
tip of $r$ falls inside phase 1 and to 0 elsewhere. The definition of 
$\rho_2(\br)$ is perfectly similar.] The scattering density of the infinitely 
large sample takes now the form: $n(\br)=n_1\rho_1(\br)+n_2\rho_2(\br)$. 
The volume fraction $\varphi_1$ of phase 1 is given 
by the relation $\varphi_1=\lim_{V\to\infty}(\int_V\rho_1(\br)\rdv/V)$ and 
is, therefore, equal to the mean value of $\rho_1(\br)$. The 
volume  fraction $\varphi_2$ of phase 2 is similarly defined. Since 
$\rho_1(\br)+\rho_2(\br)\equiv 1$, one  
has $\varphi_1+\varphi_2=1$. %This relation, combined with the positiveness 
%of the $\rho_i(\br)$s, implies that $0\le\varphi_i\le 1$ with $i=1,2$. 
By the above relations one immediately finds that the mean scattering density 
value, relevant to the infinitely large sample,  is $\barn=n_1\varphi_1+n_2\varphi_2$. 
The scattering density fluctuation takes the form
\begin{eqnarray}\label{1.12}
&&\eta(\br)=
%% n_1\rho_1(\br)+n_2\rho_2(\br)-\barn(\rho_1(\br)+\rho_2(\br))=\\ &&
(n_1-\barn)\rho_1(\br)+(n_2-\barn)\rho_2(\br)=\\ 
&&\quad (n_1-n_2)\varphi_2\rho_1(\br)-(n_1-n_2)\varphi_1\rho_2(\br).\nonumber
\end{eqnarray}           %%% \end{document}
The existence of $\Iv(0^+)$, by the same considerations  reported above 
equation (\ref{1.10D}) and the observation that the integral of $\eta(\br)$ over 
$\Vo$ continuously depends on $V$ and $O$, requires and is ensured by the following asymptotic behavior 
\begeq\label{1.12bis}
\Big|\int_{\Vo}\eta(\br)dv\Big|\approx |(n_1-n_2)\,\rho|\,V^{1/2},
\endeq     %%%%%  \end{document}
where $\rho$ is a constant with dimensions $[L^{3/2}]$. Using (\ref{1.12}) and 
the properties  $\rho_2(\br)=1-\rho_1(\br)$ and $\varphi_1+\varphi_1=1$, the above integral converts into 
%%\end{document}
\begin{eqnarray}\nonumber
&&\Big |\int_{\Vo}\eta(\br)dv\Big|=|n_1-n_2|\Big|\Bigl[(\varphi_1+\varphi_2)
\int_{\Vo}\rho_1(\br)dv-\varphi_1\,V\Bigr]\Big|=\\
&&\quad\quad\quad\quad\quad|n_1-n_2|\Big|
\int_{\Vo}(\rho_1(\br)-\varphi_1)dv\Big|.\label{1.12ter}
\end{eqnarray}
The comparison of (\ref{1.12ter}) to (\ref{1.12bis}) yields 
\begin{eqnarray}\label{1.13A} 
\int_{\Vo} \rho_1(\br)\rdv &\approx& \varphi_1\,V+\rho\,V^{1/2},%%\\
%%\int_V \rho_2(\br)\rdv &\approx& \varphi_2\,V-\rho\,V^{1/2},\label{1.13B}
\end{eqnarray}
because the absolute value can be omitted proceeding as in appendix A. 
Then, equation (\ref{1.12bis}) can be recast in the form 
%where $\rho$ is a constant with dimensions $[L^{3/2}]$ and the opposite signs 
%of the sub-leading asymptotic terms in (\ref{1.13A}) and (\ref{1.13B}) are a 
%consequence of the fact that the sum of the integrals is identically equal to $V$. 
%Using  (\ref{1.13A}), (\ref{1.13B}) and  (\ref{1.12}) one finds that the integral 
%of the scattering density fluctuation of a two homogenous phase sample must 
%asymptotically behave as
\begeq\label{1.14}
\int_V\eta(\br)\rdv \approx  (n_1-n_2)\rho\,V^{1/2}.
\endeq 
 If condition (\ref{1.14}) is obeyed, one finds that 
\begeq\label{1.15}
\Iv(0^+)=\nu^2=(n_1-n_2)^2\,v_0\quad {\rm with}\quad v_0\equiv\rho^2.
\endeq
Since $v_0$ has the dimension of a volume, the above relation 
shows that the $\bq\to 0$ limit of the observed scattering intensity is equal 
to the phase contrast times a typical volume that, in turns, is the 
square value of the coefficient of the $O(V^{1/2})$ term in the 
asymptotic expansion of the integral of the characteristic function 
of one of the sample phases. \\ 
We have already emphasized that the existence of  $\Iv(0^+)$ 
constraints $\eta(\br)$ to be such that its integral over $V$ asymptotically 
behaves as reported in equation (\ref{1.10C}) or as in equation (\ref{1.14}) 
in the case of  two homogenous phase samples. In the last case, the constraint 
can further more be elaborated. To this aim, generalizing the approach 
of M\'ering and Tchoubar (1968), we partition the infinitely 
large sample into a sequence of nested hollow spheres  $\cS_i$  
(with $ i=1,2,\ldots$) of equal volume $V_0$. 
We denote by $R_i$ the inner radius of  $\cS_i$ and, 
of course, we set  $R_1=0$ because $\cS_1$ is a sphere.  The thickness 
$\delta_i$ of $\cS_i$ is given by $\delta_i=(3V_0/4\pi+R_i^3)^{1/3}-R_i$ 
that, for $i=1$, yields $\delta_1=(3V_0/4\pi)^{1/3}$.  Since $R_2=\delta_1$ 
and $R_i=R_{i-1}+\delta_i$ if $i\ge 2$, one can recursively 
determine all the $R_i$s and $\delta_i$s. %\end{document}
We denote now  by $\rv_i$ the volume of the portion of $\cS_i$ that is occupied 
by phase 1. Then, if $V$ is taken equal to the set occupied by the first $N$ 
hollow spheres, one finds that 
\begeq\label{1.16}
\int_V\rho_1(\br)\rdv = N\bigl(\frac{1}{N}\sum_{i=1}^N \rv_i\bigr)
\endeq
We denote by $\bar \rv$ the limit of the arithmetic mean present 
within the brackets on the rhs of 
(\ref{1.16}) as $N\to\infty$.  Recalling (\ref{1.13A}), it clealry results that 
${\bar v}=\varphi_1 V_0$. Then, adding and subtracting ${\bar \rv}$ 
to each $\rv_i$ on the rhs of (\ref{1.16}) and setting $\xi_i\equiv (\rv_i-{\bar\rv})$, 
the equation converts into 
\begeq\label{1.17}
\int_V\rho_1(\br)\rdv =
%NV_0{\bar v} +N\bigl(\frac{1}{N}\sum_{i=1}^N\xi_i\bigr)=
\varphi_1\,V+\sum_{i=1}^N\xi_i.
\endeq
The comparison of this relation to  (\ref{1.13A}) shows that the  sum on the 
rhs must behave as 
$\rho V^{1/2}=\rho(NV_0)^{1/2}$, \ie
\begeq\label{1.17A}
\sum_{i=1}^N\xi_i \approx \rho{V_0}^{1/2}\,N^{1/2}.
\endeq
 Squaring one finds 
\begeq\label{1.18}
\Big[\sum_{i=1}^N\xi_i\Big]^2=\sum_{i=1}^N\xi_i^2+
2\sum_{1\le i<j\le N} \xi_i\xi_j\approx \rho^2 V_0\,N.
\endeq 
The $\xi_i$s can be looked at as a  sequence of random numbers with arithmetic 
mean value equal to zero. Consequently, the sum involving the $\xi_i^2$s in 
the middle of ({\ref{1.18}), once it is divide by $N$, yields the variance of the 
random sequence in the limit $N\to \infty$. We  assume that this variance is 
finite and we denote its value by  $\mu^2V_0$. This assumption amounts 
to asymptotically write 
\begeq\label{1.19}
\sum_{i=1}^N\xi_i^2 \approx \mu^2\,V_0\,N.
\endeq 
Then,  the validity of (\ref{1.18}) requires that 
\begeq\label{1.20}
\sum_{1\le i<j\le N} \xi_i\xi_j\approx \mu'\,V_0\,N\quad{\rm with}\quad  \mu'\equiv\rho^2-\mu^2.
\endeq 
The  sum present in (\ref{1.20})  involves $N(N-1)/2$ addends. Thus, while in 
the case of (\ref{1.19}) it is sufficient to assume that the ${\xi_i}^2$s have a 
finite upper bound for the equation to be true, to work out the constraints 
that make equation (\ref{1.20}) valid is not simple. 
In appendix B we report some examples of random sequences that respectively 
obey none of  (\ref{1.19}) and (\ref{1.20}) or one of these or both. This result 
further  confirms the conclusion that:  a sequence of $\rv_i$s is physically 
realizable [\ie\,the $\rv_i$s form the volume sequence of one of the two 
homogeneous phases of a real sample] if  the associated $\xi_i$s obey both  
(\ref{1.19}) and (\ref{1.20})  because only in this case the $\Iv(0^+)$ exists 
and is intensive with respect to $V$. In conclusion,  we can state:  (C) 
{\em the sum $\sum_{1=1}^N{\xi_i}$ 
asymptotically behaves as $N^{1/2}$, times a constant of dimensions 
$[{\rm L}^{3/2}]$, if the random sequence of the $\xi_i$s has mean value equal 
to zero, finite variance and obeys (\ref{1.20}). The intensity limit value, 
in terms of the last quantities, reads}
\begeq\label{1.21}
\Iv(0^+)\approx I(0^+)=(n_1-n_2)^2(\mu^2+\mu')=(n_1-n_2)^2\rho^2. 
\endeq  
We also add the followingl remarks:\\
i)   statement C) is similar to the 
central limit theorem (Rosenthal,  2000). This theorem states that the 
sum of $N$ independent and identically distributed random variables 
asymptotically behaves, in distribution, as $N^{1/2}$ times the normal 
distribution. Hence, in comparison to the central limit theorem, 
statement C) substitutes the convergence in distribution with the asymptotic 
convergence  and the assumption of independent and identically distributed random variables with  conditions (\ref{1.19}) and (\ref{1.20});\\ 
ii) in deriving (\ref{1.21}) no bounds  on the particle size, shape and 
polidispersity were required. Assuming the sample made up of a single 
kind of particles,  it is possible to relate  
the $\Iv(0^+)$ value [Guinier \& Fournet (1955), Hansen \& McDonald (1976), 
Luzzati (1995)] to the isothermal compressibility of the sample. The 
last quantity is related to the mean square fluctuation of the particle 
number  (Landau \& Lifshitz, 1967, Sect. 114). Interestingly this result can 
simply be obtained by slightly changing the procedure expounded above 
equation (\ref{1.16}) to account for the hypothesis that each particle 
is rigid and has volume $v_p$. To this aim, it is first observed that 
each  particle  has its gravity center  inside 
one and only one of the $\cS_i$s.  Then the (outer) border of $\cS_1$ is 
as slightly as possible  modified so as the new  ${\cS_1}'$ has still volume 
$V_0$, fully contains all the particles having  their gravity centers  
lying within $\cS_1$ and fully excludes those with their centers lying outside 
$\cS_1$.  We denote by ${N_1}'$ the number of particles  
present in $\cS_1'$. The substitution of $\cS_1$ with ${\cS_1}'$ will 
clearly require the change of $\cS_2$ into ${\cS_2}'$. The inner border of 
${\cS_2}'$  is the border of ${\cS_1}'$. The outer border 
is fixed by the conditions that ${\cS_2}'$ has volume $V_0$ and fully 
contains all the  (and  only the) particles that have their gravity centers lying  
inside $\cS_2$.  The relevant particle number will be denoted by ${N_2}'$. 
In this way one determines, step by step, ${\cS_3}'$, ${\cS_4}'$ and so 
on. Provided $V_0$ be not too small and the density of the system 
not too high, the procedure ought to work. Assuming  
this point  fully proved, we pass now to evaluate the left hand 
side of (\ref{1.14}).  Using (\ref{1.12}) and the property that 
$\varphi_1+\varphi_2=1$ one finds 
\begin{eqnarray}
\ & &\int_V\eta(\br)\rdv=(n_1-n_2)\bigl[{\sum_{i=1}^N} v_p\varphi_2 {N_i}'-
\varphi_1\big(NV_0-{\sum_{i=1}^N} v_p{N_i}'\bigr)\bigr]=\nonumber\\
&&\quad\quad\quad\quad  (n_1-n_2)v_p[\cN_{V,0}-\varphi_1\,N V_0/v_p],\label{1.22}
\end{eqnarray}
where  $\cN_{V,0}\equiv \sum_{i=1}^N {N_i}'$ represents the number of 
particles contained within $V$ while ${\bar N}_0\equiv \varphi_1\,N V_0/v_p$ represents the mean number of particles contained within $V$ as this  becomes infinitely large.  [It is noted that one needs to know $\varphi_1$ in order to know ${\bar N}_0$.] Subsript 0 recalls that both $\cN_{V,0}$  and ${\bar N}_0$ values depend on the $V_0$ choice.  By (\ref{1.22}) it follows that 
\begin{eqnarray}
&& 
\frac{1}{V}\Bigl[\int_V\eta(\br)\rdv\Bigr]^2=
(n_1-n_2)^2v_p^2\frac{[\cN_{V,0}-\varphi_1\,N V_0/v_p]^2}{NV_0}\approx \nonumber\\
&& \quad \quad\quad\quad  (n_1-n_2)^2v_p^2 \lim_{N\to\infty}\frac{(\cN_{V,0}-{\bar N}_0)^2}{NV_0}.\label{1.23}
\end{eqnarray}
By the same argument used above equation (\ref{1.10C}), one has that 
$(\cN_{V,0}-{\bar N}_0)=O(N^{1/2})$. and one can therefore write 
%(\ref{1.23}) as 
%\begeq\label{1.24}
%\frac{1}{V}\Big[\int_V\eta(\br)\rdv\Big]^2\approx \frac{ (n_1-n_2)^2v_p^2}%{v}\overline{(\cN-{\bar N})^2}
%\endeq
\begeq\label{1.24}
\lim_{N\to\infty}\frac{(\cN_{V,0}-{\bar N}_0)^2}{NV_0}=\frac{1}{v}
\endeq
where, by dimensional analysis, $v$ is a typical volume that must be 
independent on $V_0$ for consistency. [This implies that the numerator 
in the left hand side of (\ref{1.24}) is linear in $V_0$.]  The left hand side 
of  (\ref{1.24}) can be looked at as  a procedure able to evaluate the mean 
particle number fluctuation in the case of  physical samples made of fixed 
and rigid particles of the same volume but not necessarily of the same 
shape, provided $\varphi_1$ be known.  \\ 
iii) statement C) can simply be applied to an ideal simple cubic crystal to 
conclude that $\Iv(0^+)=0$. In fact, one  approximates the atoms at the 
centers of  the cells by  hard bodies of fixed shape  and volume $v_0$.
Denoting the cell size by $a$, the unit cell  volume is $V_c\equiv a^3$. 
The sequence of cubes $V_k$, having the same  gravity center and 
orientation, and size edges equal to $(2k+1)a$, is such that the limit 
of $V_k$ as $k\to\infty$ is equal  to the volume of the infinitely large 
sample.  We name phase 1 that formed by the hard bodies. Then,  
$\varphi_1=v_0/V_c$ and  equation (\ref{1.16}) yields 
\begeq
\frac{1}{V_k}\int_{V_k}\rho_1(\br)\rdv = \frac{(2k+1)^3 v_0}{(2k+1)^3 V_c}=\varphi_1. 
\endeq
The rhs does not depend on $k$ and one finds that 
$\int_V\rho_1(\br)\rdv/V= \varphi_1$ which shows that no $O(V^{1/2})$ contribution is present. Thus, $\rho=0$ and, consequently, $\Iv(0^+)$ is equal to zero for simple cubic crystals. We refer to Gommes (2016) for further geometries characterized by vanishing $I_V(0^+)$ values.
%\vfill\eject
%%%%%%%%%%%%      CONCLUSION
\subsubsection*{Conclusions}
Since physical systems obey properties A) and B), a physical scattering density 
fluctuation must asymptotically behave according to equation (\ref{1.10C}) or 
to (\ref{1.12bis}) and (\ref{1.13A}) in the case of samples made up of two homogeneous phases. This result can be put in a form similar to Porod's law 
in the sense that the plot of $\Bigl(V^{1/2}\,\int\eta_V(\br)\rdv\bigr)$ 
{\em versus} $V$  
shows a plateau of  height $\nu$ (both positive and negative) at 
large $V$s. The practical application of this procedure is however much 
more  ambiguous than in Porod's cases (Ciccariello {\em et al.},1988) unless $\eta(\br)$ is 
analytically known, a very exceptional case indeed. In most of the cases,   
$\eta(\br)$ is generated by numerical simulations over a rather small 
spatial domain and, consequently, the application of the above recipe 
does not yield fully consistent results. The three panels of Fig.1, reported for greater completeness, illustrate these aspects. They refer  to the simplest case 
of N random points $x_i$ uniformly generated within the interval $[0,\,1]$ with the further constraint that their relative distance are greater than $\sigma\equiv\varphi_1/N$. We set $\varphi_1=0.2$ and considered the 
cases: $N=10^3,\, 10^4$ and $10^5$.  Since the value of $\sigma$ 
decreases  as $N$ increases, the onset of  the asymptotic 
behavior  ought be more evident in the case $N=10^5$, as it really happens. 
For this reason, the shown panels refer to $N=10^5$.  In the upper panel, we
have interpreted the segments $[0,x_1],\,[x_2,x_3],\,[x_4,x_5],\ldots$ and $[x_1,x_2]$, $[x_3,x_4],\ldots$ as those respectively 
relevant to phases 1 and 2.  Starting from the origin  we evaluated the 
mean value ${\bar\rho}_L$ of $\rho_1(x)$ over the interval of length $L$. 
Fitting the resulting values  to the function $\varphi+\rho_0\,L^{1/2}$ 
in the range $0.5<L<1$, we determined both $\varphi$ and $\rho_0$.  
In particular the resulting value  $\varphi=0.5004$ looks quite accurate 
owing to the uniform distribution of the $x_i$s. 
The panel shows the plot of the resulting $L^{1/2}\bigl({\bar\rho}_L-\varphi\bigr)$ quantity. The approach 
to a plateau  appears evident. 
The lower two panels refer to a different model obtained by the generated 
$x_i$s, since each of these points is interpreted as the center of an interval 
of length $\sigma$. These intervals form phase 1 and the complement of 
their union with respect to interval $[0.\,1]$ phase 2.  
\begin{figure}[hp]
\includegraphics[width=7.truecm]{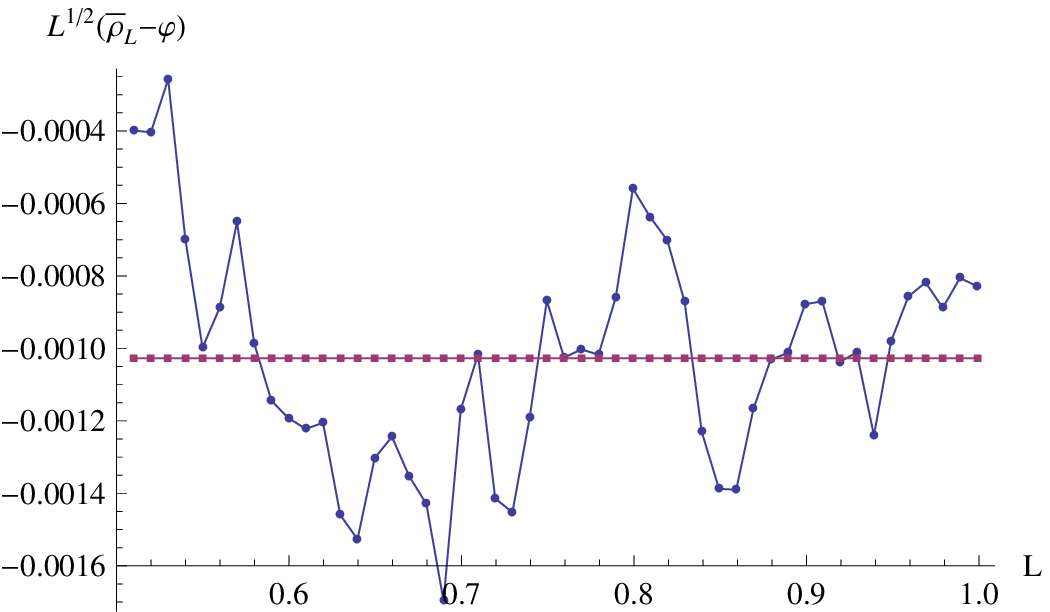}

\includegraphics[width=7.truecm]{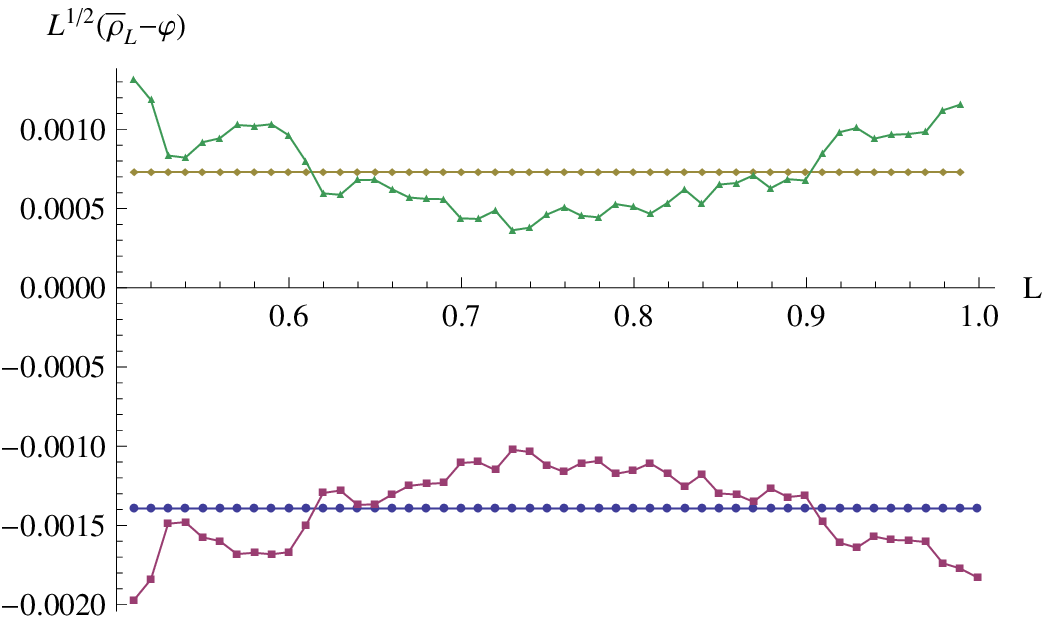} 
\includegraphics[width=7.truecm]{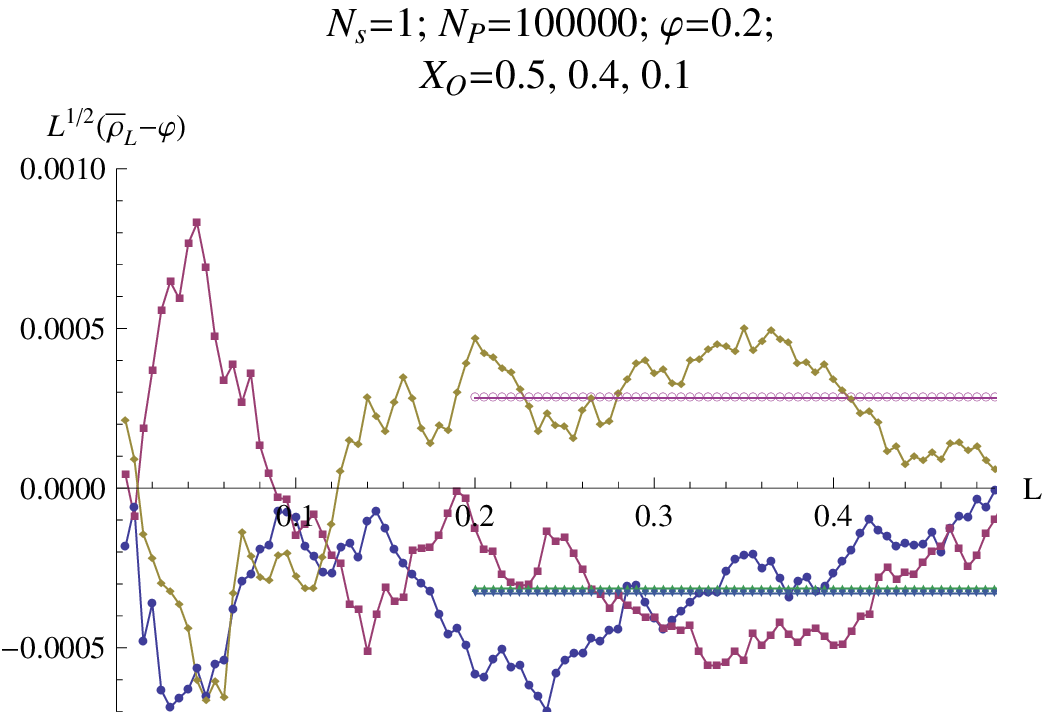}
\caption{\label{Fig1} {The three panels numerically illustrate the approach of 
one-dimensional random generated scattering densities to the theoretical behavior predicted by equations (\ref{1.12bis}) and (\ref{1.13A}). }}
\end{figure}
The bottom left panel 
shows the behavior of the fitted $L^{1/2}\bigl({\bar\rho}_L-\varphi\bigr)$ 
quantities for phase 1  (bottom red curve)  and phase 2 
 (top green curve). In the two cases the values of $\varphi$, resulting by the 
fits, are respectively 0.1998 and 0.7995, quite close to the exact 0.2 and 0.8 values.  The corresponding $\rho_0$ values are the plateau heights. They 
ought to be opposite while they are $-0.00138$ and $ 0.00086$ and thus 
the condition is only weakly obeyed. The last panel (bottom right) checks 
the independence of the mean values if one chooses different origins. 
The blue, red and golden curves refer to origins sets at $X_0= 0.5,\, 0.4$ and $0.1$. To attain the value $L=1/2$ in the three cases, the $x_i$ sequence 
was periodically replicated on the left and on the right. In the nearly 
asymptotic range $L>0.2$, the plateau heights were determined by the 
the expounded  best-fit. In contrast to the choice $X_0=0.1$, going from $X_0=0.5$ to $X_0=0.4$ can be considered a small shift since the plateau does not appreciably change.  This conclusion is not unexpected when the 
scattering density is generated on a finite interval. Overall, the shown cases 
confirm the difficulty in numerically applying relations (\ref{1.10C}) or 
to (\ref{1.12bis}) and (\ref{1.13A}),  though the usefulness of the 
relations to better characterize physical scattering density cannot be denied   on a theoretical ground. 
%%%%%%%%%          ACKNOWLEDGMENTS     %%%%%%%%%%%%%%%
\subsubsection*{Acknowledgmets} 
I   thank Prof. C. Gommes for a criticism on the first draft of this paper,  
Prof.s  K. Lechner and  P.A. Marchetti for a useful discussion and 
Prof.s A. Benedetti, W. Gille and P. Riello for useful conversations and correspondence. 
%%%%%%%%%%%%%        APPENDIX  A               %%%%%%%%%%%%%%%
\subsubsection*{Appendix A: proof of equation (\ref{1.10C})}
%%%%%%%%%%%%%%  
Equation (\ref{1.10Ca}) is equivalent to 
\begeq\label{A.1}
\lim_{V\to\infty}\Big(\Big|\int \eta_{\Vo}(\br)\rdv\Big|/V^{1/2} \Big)= |\nu|, 
\endeq
Let us first  keep $O$ fixed. The limit mathematical definition ensures that, 
for any  $\epsilon>0$,   implies that the absolute  value of the difference of 
the two sides of (\ref{A.1}) is smaller than $\epsilon$ if  $V>V_{\epsilon}$. 
Choosing $\epsilon$ in such a way that $\epsilon<|\nu|$, one concludes that
the left hand side of (\ref{A.1}) is positive for all the  sets $V_0$ 
of volume greater than $V_{\epsilon}$. Consequently for all the $\Vo$s such 
that $V>V_{\epsilon}$ the sign of $\int \eta_{\Vo}(\br)\rdv$, that continuously depends on $\Vo$,  is either positive or negative because  the integral never 
vanishes. One can therefore write %\end{document}
\begeq\label{A.2}
\int \eta_{\Vo}(\br)\rdv\approx \nu_o\,V^{1/2} ,\quad
{\rm with}\quad |\nu_o|=|\nu|.
\endeq
The assumption of a fixed $O$ is now removed. Consider  a different origin $O\p$.  The above reasoning holds true provided $\nu_o$ is substituted 
with $\nu_{o\p}$. The absolute values of these two constants are equal. 
Hence, either $\nu_{o\p}=\nu_{o}$ or $\nu_{o\p}=-\nu_{o}$.  
Let $O$ span  all the space. This divides into two regions. The first is 
formed by all the $O\p$ such that $\nu_{o\p}=\nu_{o}$ and the second 
by the $O\p$ such that $\nu_{o\prime}=-\nu_{o}$. Consider now two 
origins $O_1$ and $O_2$, very close to each other and respectively lying within 
the first and second region.  Let $O$ denote the center of a set $\Vo$  and let $O$ continuously move from $O_1$ to $O_2$.  Besides let the $\Vo$'s volume be   
so large that the integral obeys to its asymptotic behavior. The integral
continuously depends on the $O$ position. This property is clearly contradicted 
by the fact the the asymptotic leading term takes opposite values as $O$ goes from 
$O_1$ to $O_2$. This proves that equality $\nu_{o\p}=-\nu_{o}$ cannot occur 
and eqaution (\ref{1.10C}) is proved. \\ 
We conclude this section reporting an example of scattering density fluctuation that 
obeys condition (\ref{1.10C}). Consider first the one dimensional case and the 
function 
\begeq\label{A.3}
F_{\eta}(x)\equiv \sin^2(x)/|x|^{1/2}.
\endeq
By MATHEMATICA software (Wolfram Research, Champaign, IL, USA) one finds that 
\begeq\label{A.4}
\int_{a}^{a+L}F_{\eta}dx=(a+L)^{1/2}-a^{1/2} + (\pi/2) \Bigl(\cC(2(a/\pi)^{1/2})
-\cC(2((a+L)/\pi)^{1/2})\Bigr),
\endeq 
where $\cC(\cdot)$ is the cosine Fresnel integral (Abramowitz \& Stegun, 1970). 
Its leading asymptotic expansion at large $L$  simply reads 
\begeq\label{A.5}
\int_{a}^{a+L}F_{\eta}dx =  L^{1/2}\Bigl(1+\big[\pi^{1/2}\cC(2(a/\pi)^{1/2}))/2-a^{1/2}-\pi^{1/2}/4\bigr]/L^{1/2}\Bigr) +o
\endeq
and it agrees with the one dimensional version of (\ref{1.10C}). By this result 
it is trivial to show that the function $F_{\eta}(x)F_{\eta}(y)F_{\eta}(z)$ obeys 
(\ref{1.10C}) and, therefore, represents a candidate for a physical scattering density 
fluctuation.
%%%%%%%%%%%%%        APPENDIX  B  
\subsubsection*{Appendix B: an example of random sequence obeyng (\ref{1.19}) and (\ref{1.20})}  
%%%%%%%%%%%%%        APPENDIX B   
We explicitly show that both condition (\ref{1.19}) and (\ref{1.20}) must be fulfilled for (\ref{1.17A}) to be fulfilled . To this aim, we assume that the $\xi_i$s, defined below equation (\ref{1.16}), have the form
\begeq\label{B.1}
\xi_{3\jmath-2}=a\,\jmath^{s}+\jmath^{-r},\quad  \xi_{3\jmath-1}=-a\,\jmath^{s}+\jmath^{-r}\quad {\rm and}\quad
\xi_{3\jmath}=\jmath^{-r},
\endeq 
with $r>0$, $-1<s<1$ and $\jmath=1,2,\ldots$. 
Put
\begin{eqnarray}\label{B.2}
S_{3N}&\equiv&\sum_{\jmath=1}^N\bigl(\xi_{3\jmath-2}+\xi_{3\jmath-1}+
\xi_{3\jmath}\bigr)\\
{\rm Var}_{3N}&\equiv&\sum_{\jmath=1}^N\bigl({\xi_{3\jmath-2}}^2+
{\xi_{3\jmath-1}}^2+{\xi_{3\jmath}}^2\bigr)/{3N}.
\end{eqnarray}
MATHEMATICA  yields %\end{document}
\begeq\label{B.4}
S_{3N}=3\,{H_N}^{(r)}\quad {\rm and}\quad 
{\rm Var}_{3N}=\Bigl(3\,{H_N}^{(2r)}+2\,a^2\,{H_N}^{(-2s)}\Bigr)/3N,
\endeq 
where ${H_N}^{(r)}$ denotes the generalized harmonic number function 
(Erhardt, 2016). 
The leading asymptotic expansions,  with respect to $N$, of $S_{3N}$ and 
${\rm Var}_{3N}$ are 
\begeq\label{B.5}
S_{3N}\approx 3\,N^{r}\,\zeta(r)+3\,N^{1-r}/(1-r)+3/(2N^r)
\endeq
and        %    \end{document}
\begeq\label{B.6}
{\rm Var}_{3N}\approx 
\begin{cases}
\frac{2 a^2 N^{2s}}{6s+3}+
\frac{\zeta(2r)}{N}+
\frac{2a^2\zeta(-2s)}{3N}+\frac{1}{(1-2r) N^{2r} }  &\text{if}\  r\ne 1/2,\\
 \frac{2a^2N^{2s}}{3(1+2s)}+\frac{3\gamma_C+\log(N)+2a^2\zeta(-2s)}{3N}    &\text{if}\  r=1/2.
\end{cases} 
\endeq
where $\zeta(\cdot)$ denotes the Riemann zeta function and $\gamma_C$ 
the Euler-Mascheroni constant. Equation (\ref{B.5}) shows that, whatever $s$,  $S_{3N}$ behaves as $\sqrt{N}$ if and only if $r=1/2$ while, for different 
$r$s, it increases faster.   Equation (\ref{B.6}) shows that ${\rm Var}_{3N}$ 
diverges with $N$  if $s>0$. Hence, if $s>0$ and $r=1/2$, the left hand 
side of  (\ref{1.20}) also must diverge to cancel the variance divergence 
because (\ref{1.17A}) is $O(N^{1/2})$. If $s=0$, the variance is finite and, 
therefore, sum (\ref{1.20}) diverges faster than $\sqrt{N}$ if $r\ne 1/2$ 
and exactly behaves as $\sqrt{N}$ if $r=1/2$.  The above conclusions also 
apply to $S_{3N+i}$ and ${\rm Var}_{3N+i}$ with $i=1,2$ and are, therefore, 
fully general. 
%\end{document}
\vfill\eject
\section*{References}
\begin{description}
\item[\refup{}]Abramowitz, M. \& Stegun, I.A. (1970). {\em Handbook of Mathematical Functions}, New York: Dover.
\item[\refup{}]  Ciccariello, S. (2002). {\em Acta Cryst.} A58, 460-463.
\item[\refup{}] Ciccariello, S., Goodisman, J. \& Brumberger, H. (1988). 
 {\em  J. Appl. Cryst.}  {\bf 21}, 117-128
%\item[\refup{}]  Ciccariello, S. (2016). {\em In preparation}. 
%\item[\refup{}] Debye, P., Anderson, H.R.  \&  Brumberger, H. (1957). {\em J. Appl.
%Phys.} {\bf 20}, 679-683.
\item[\refup{}] Debye, P., Anderson, H.R.  \&  Brumberger, H. (1957). {\em J. Appl.
Phys.} {\bf 20}, 679-683.
\item[\refup{}] Erhardt, W. (2016). {\em 
http://www.wolfgangehrhardt.de/specialfunctions.pdf}.
\item[\refup{}] Feigin, L.A. \&  Svergun, D.I. (1987). {\em Structure Analysis
by Small-Angle X-Ray and  Neutron Scattering}, New York: Plenum Press.
\item[\refup{}] Gommes, C. (2016). {\em J. Appl. Cryst.} {\bf 49}, 1162-1176.
%\item[\refup{}]   Gille, W. (1999). {\em J. Appl. Cryst.} {\bf 32}, 1100-1104.
%\item[\refup{}]   Gille, W. (2014). {\em Particle and Particle Systems
% characterization }, London: CRC.
%\item[\refup{}] Glatter, O. (1982). {\em Small-Angle X-Ray Scattering}. 
% Edts Glatter, O. \& Kratky, O., London: Academic Press. 
\item[\refup{}] Guinier, A. \& Fournet, G. (1955). {\em Small-Angle Scattering of X-rays.} New York: John Wiley.
\item[\refup{}] Hansen, J. P. \& McDonald, I. R. (1976). {\em Theory of Simple Liquids}, Section 4.2. London: Academic Press.
\item[\refup{}] Kostorz, G. (1979). {\em Neutron Scattering}, Ed.  Kostorz, G.,   
London: Academic Press,  pp 227-289.
\item[\refup{}] Landau, L.D. \& Lifshitz, E. (1967a). {\em Physique Statistique}.  Moscou: \'Editions MIR.
\item[\refup{}] Landau, L.D. \& Lifshitz, E. (1967b). {\em  Physique des Milieux Continues}.  Moscou: \'Editions MIR.
\item[\refup{}] Lantuejoul, Ch.  (1990). {\em Ergodicit\'e et porte\'ee inte\'grale}, 
Ecole de Mines de Paris, Report 17/90/G  .
\item[\refup{}] Luzzatti, V. (1995). {\em Modern Aspects of Small-Angle Scattering.}
Edt. Brumberger, H., Dordrecht: Kluivert Acad. Pubs.
\item[\refup{}] M\'ering, J. \& Tchoubar, D. (1968). {\em J. Appl. Cryst.} {\bf 1}, 153-65.
\item[\refup{}] Rosenthal, J.S. (2000). {\em A first look at rigorous probability theory}. 
Singapore: World Scientific.
%\item[\refup{}] https://en.wikipedia.org/wiki/Central\_limit\_theorem

\end{description}
\end{document}